\documentclass[a4,12pt]{article}

\usepackage{graphicx}

\begin{document}

\begin{center}
{\large\bf A Possible Detection of the Cosmic Antineutrino Background \\
in the Presence of Flavor Effects}
\end{center}

\vspace{0.3cm}
\begin{center}
{\bf Y.F. Li} \footnote{E-mail: liyufeng@ihep.ac.cn}
~ and ~ {\bf Zhi-zhong Xing}
\footnote{E-mail: xingzz@ihep.ac.cn} \\
{Institute of High Energy Physics $\&$ Theoretical Physics
Center for Science Facilities, \\
Chinese Academy of Sciences, Beijing 100049, China}
\end{center}

\setcounter{footnote}{0}

\vspace{2.5cm}

\begin{abstract}
Lusignoli and Vignati have recently pointed out that it is in
principle possible to directly detect the cosmic {\it antineutrino}
background (C$\overline{\nu}$B) by using the rather stable isotope
$^{163}{\rm Ho}$ as a target, which can decay into $^{163}{\rm Dy}$
via electron capture (EC) with a very small energy release. In this
paper we calculate the rate of the relic antineutrino capture on
$^{163}{\rm Ho}$ nuclei against the corresponding EC decay rate by
taking account of different neutrino mass hierarchies and reasonable
values of $\theta^{}_{13}$. We show that such flavor effects are
appreciable and even important in some cases, and stress that a
calorimetric measurement of the C$\overline{\nu}$B might
be feasible in the far future. \\

\hspace{-0.7cm}
\end{abstract}

\newpage

The standard cosmology predicts that relic neutrinos and
antineutrinos of the Big Bang must survive today and form a cosmic
background, but their average temperature is so low ($T^{}_\nu
\simeq 1.945$ K) that their existence has not directly been verified
\cite{XZ}. Among several proposals for a direct detection of the
cosmic {\it neutrino} background (C$\nu$B) \cite{Ringwald}, the most
promising one should be the neutrino capture experiment by means of
radioactive $\beta$-decaying nuclei
\cite{Weinberg}--\cite{Faessler}. A direct measurement of the cosmic
{\it antineutrino} background (C$\overline{\nu}$B) could similarly
be done by using radioactive nuclei which can decay via electron
capture (EC) \cite{Cocco2}, but it seems much less hopeful.

Lusignoli and Vignati have recently pointed out an interesting and
noteworthy way to directly detect the C$\overline{\nu}$B
\cite{Lusignoli}. It takes advantage of the rather stable isotope
$^{163}{\rm Ho}$ as a target, which undergoes an EC decay into
$^{163}{\rm Dy}$ with a small energy release ($Q \simeq 2.5$ keV).
The signature of this relic antineutrino capture process is located
on the right-hand side of the spectral endpoint of the EC decay, and
their interval is detectable if the experiment has a sufficiently
good energy resolution. In practice, it is promising to do a
high-statistics {\it calorimetric} experiment to measure the
$^{163}{\rm Ho}$ spectrum, probe the absolute neutrino mass scale
\cite{Mare} and even detect the C$\overline{\nu}$B.

In this paper we shall calculate the rate of the relic antineutrino
capture on $^{163}{\rm Ho}$ nuclei against the corresponding EC
decay rate by taking account of different neutrino mass hierarchies
and reasonable values of the smallest neutrino mixing angle
$\theta^{}_{13}$. Such flavor effects, which were not considered in
Ref. \cite{Lusignoli}, are found to be appreciable and even
important in some cases. We shall first look at the energy spectrum
of the EC decay of $^{163}{\rm Ho}$ by examining how sensitive the
ratios of the peak heights and the fine structure near the spectral
endpoint are to the neutrino mass hierarchy, and then present a
detailed analysis of flavor effects on the detection of the
C$\overline{\nu}$B by using a target made of this isotope. Although
our numerical results are mainly for the purpose of illustration, we
stress that such a direct laboratory measurement of the
C$\overline{\nu}$B deserves further
attention and investigations and might be feasible in the long term. \\

The key point of the relic antineutrino capture on $^{163}{\rm Ho}$
nuclei is to capture the relic {\it electron} antineutrinos in a way
without any energy threshold. In the three-flavor scheme the
$|\nu^{}_e\rangle$ and $|\overline{\nu}^{}_e\rangle$ states are
superpositions of three neutrino and antineutrino mass eigenstates,
respectively:
\begin{eqnarray}
\left|\nu^{}_e \right\rangle \hspace{-0.2cm} & = & \hspace{-0.2cm}
\sum_k V^*_{e k} \left|\nu^{}_k \right\rangle \; ,
\nonumber \\
\left|\overline{\nu}^{}_e \right\rangle \hspace{-0.2cm} & = &
\hspace{-0.2cm} \sum_k V^{}_{e k} \left| {\overline{\nu}^{}_k}
\right\rangle \; ,
\end{eqnarray}
where $V^{}_{e k}$ (for $k=1,2,3$) denotes an element in the first
row of the $3\times 3$ neutrino mixing matrix $V$. We have $|V^{}_{e
1}| = \cos\theta^{}_{12} \cos\theta^{}_{13}$, $|V^{}_{e 2}| =
\sin\theta^{}_{12} \cos\theta^{}_{13}$ and $|V^{}_{e 3}| =
\sin\theta^{}_{13}$ in the standard parametrization of $V$
\cite{PDG}, where $\theta^{}_{12} \simeq 34^\circ$ and
$\theta^{}_{13} < 12^\circ$ as indicated by a global analysis of
current neutrino oscillation data \cite{GG}. Of course,
$\theta^{}_{13} = 0^\circ$ and $\theta^{}_{13} \neq 0^\circ$ may
have very different consequences in neutrino phenomenology, as we
shall see later on.

Now let us consider the EC decay of the isotope $^{163}{\rm Ho}$
\cite{Bambynek}:
\begin{equation}
^{163}{\rm Ho}  + e^-_{i (\rm shell)} \to  \; ^{163}{\rm Dy}^{*}_{i}
+ \nu^{}_e  \to  \; ^{163}{\rm Dy} + E^{}_i + \nu^{}_e \; ,
\end{equation}
where $e^-_{i \rm (shell)}$ denotes an orbital electron from the
$i$-th shell of holmium-163, and $E^{}_i$ is the corresponding
binding energy of the electron hole in dysprosium-163. If the
$Q$-value of this EC decay is defined as the mass difference between
the two atoms in their ground states, then the energy spectrum of
the outgoing neutrinos will be given by a series of lines at
$Q-E^{}_i$. Note that $Q$ has not well been determined and its value
is possible to range from 2.3 keV to 2.8 keV \cite{Lusignoli}. In
what follows we shall take $Q = 2.5$ keV as a default value unless
otherwise explicitly mentioned. Taking account of the Breit-Wigner
resonance form of the atomic levels \cite{Bambynek}, one may express
the energy spectrum of the detected EC events as follows:
\begin{eqnarray}
{{\rm d} \lambda^{}_{\rm EC}\over {\rm d}T^{}_{\rm c}}
\hspace{-0.2cm} &=& \hspace{-0.2cm} {G^{2}_{\beta} \over {4
\pi^2}}(Q-T^{}_{\rm c}) \sum_k |V^{}_{ek}|^2 \sqrt{(Q-T^{}_{\rm
c})^2-m_{k}^2}~ \Theta(Q - T^{}_{\rm c} - m^{}_k)
\nonumber \\
&& \hspace{-0.2cm} \times \sum_i n^{}_i \, C^{}_i \, \beta_i^2 \,
B^{}_i \, {\Gamma^{}_i \over 2\pi}\cdot {1 \over (T^{}_{\rm
c}-E^{}_i)^2+\Gamma_i^2/4} \; ,
\end{eqnarray}
where $\lambda^{}_{\rm EC}$ denotes the rate of this EC decay,
$T^{}_{\rm c}$ is the so-called ``calorimetric" energy which
measures the difference between $Q$ and the neutrino energy
\cite{Rujula}, $G^{}_{\beta} \equiv G^{}_{\rm F} \cos \theta^{}_{\rm
C}$ with $\theta^{}_{\rm C} \simeq 13^\circ$ being the Cabibbo angle
of quark flavor mixing, $m^{}_k$ is the mass of
$\overline{\nu}^{}_k$ (for $k=1,2,3$), and the theta function
$\Theta(Q - T^{}_{\rm c} - m^{}_k)$ has been imposed on Eq. (3) to
ensure the kinematic requirement. In addition, $n^{}_i$ denotes the
fraction of occupancy of the $i$-th electron shell, $C^{}_i$ is the
nuclear shape factor, $\beta^{}_i$ represents the Coulomb amplitude
of the electron radial wave function, and $B^{}_i$ is an atomic
correction for the electron exchange and overlap. The $Q$-value of
this EC decay is so small that only those electrons from $M^{}_1$,
$M^{}_2$, $N^{}_1$, $N^{}_2$, $O^{}_1$, $O^{}_2$ and $P^{}_1$ levels
can be captured \cite{Lusignoli}. In an excellent narrow-width
approximation \cite{Rujula} the integral of Eq. (3) yields
\begin{equation}
\lambda^{}_{\rm EC} = {G^{2}_{\beta} \over {4 \pi^2}} \sum_i \sum_k
n^{}_i \, C^{}_i \, \beta_i^2 B^{}_i \left(Q-E^{}_i \right)
|V^{}_{ek}|^2 \sqrt{(Q-E^{}_i)^2-m_k^2} \; ,
\end{equation}
where $i$ runs over all the possible electron shells of $^{163}{\rm
Ho}$ that involve its EC decay, and $k$ runs over three neutrino
mass eigenstates.

Given a positive $Q$-value, a thresholdless capture of the incoming
relic antineutrinos on the EC-decaying $^{163}{\rm Ho}$ nuclei may
happen through
\begin{equation}
\overline{\nu}^{}_e + \, ^{163}{\rm Ho} + e^-_{i (\rm shell)} \to
\; ^{163}{\rm Dy}^{*}_{i} \to  \; ^{163}{\rm Dy} + E^{}_i \; .
\end{equation}
The capture rate for each mass eigenstate $\overline{\nu}^{}_k$ (for
$k=1,2,3$) hidden in the flavor eigenstate $\overline{\nu}^{}_e$ can
be given by
\begin{equation}
\lambda^{}_{\overline{\nu}^{}_k} = {G_{\beta}^2 \over 2} \,
n^{}_{\overline{\nu}^{}_k} |V^{}_{ek}|^2 \sum_i n^{}_i \, C^{}_i  \,
\beta_i^2 \, B^{}_i \, {\Gamma^{}_i \over 2\pi} \cdot {1 \over
(E^{}_{\overline{\nu}^{}_k} + Q - E^{}_i)^2 + \Gamma_i^2/4} \; ,
\end{equation}
where $n^{}_{\overline{\nu}^{}_k}$ and $E^{}_{\overline{\nu}^{}_k}$
are the number density and energy of $\overline{\nu}^{}_k$,
respectively. The standard Big Bang cosmology predicts
$n^{}_{\nu^{}_k} = n^{}_{\overline{\nu}^{}_k} \simeq 56 ~{\rm
cm}^{-3}$ today for each species of relic neutrinos and
antineutrinos \cite{XZ}, but possible gravitational clustering of
massive $\nu^{}_k$ and $\overline{\nu}^{}_k$ around the Earth could
make their number densities much bigger \cite{Wong}. For simplicity,
we shall not consider the gravitational clustering effect in the
subsequent analysis. Note that the de-excitation energy of unstable
$^{163}{\rm Dy}^{*}_{i}$ in Eq. (5) is in principle monoenergetic
for each mass eigenstate $\overline{\nu}^{}_k$ (i.e., $T^{}_k \equiv
E^{}_{\overline{\nu}^{}_k} + Q$). Convoluted with a finite energy
resolution in practice, the ideally discrete energy lines of the
final states in Eq. (5) must spread and then form a continuous
spectrum. As usual, we consider a Gaussian energy resolution
function defined by
\begin{equation}
R(T, \, T^{}_k) = \frac{1}{\sqrt{2\pi} \,\sigma} \exp\left[-\frac{(T
- T^{}_k)^2}{2\sigma^2} \right] \; ,
\end{equation}
where $T^{}$ is the overall energy of an event detected in the
experiment. Using $\Delta$ to denote the experimental energy
resolution (i.e., the full width at half maximum of a Gaussian
energy resolution for the detected events), we have $\Delta =
2\sqrt{2\ln 2} \,\sigma \approx 2.35482 \,\sigma$. Then the
differential antineutrino capture rate reads
\begin{eqnarray}
{{\rm d} \lambda^{}_{\overline{\nu}}\over {\rm d} T^{}} =
{G_{\beta}^2 \over 2} \sum_i \sum_k n^{}_{\overline{\nu}^{}_k}
|V^{}_{ek}|^2 R(T, \, T_k) \, n^{}_i \, C^{}_i  \, \beta_i^2 \,
B^{}_i \, {\Gamma^{}_i \over 2\pi} \cdot {1 \over (T^{}_k-E^{}_i)^2
+\Gamma_i^2/4} \; .
\end{eqnarray}
In the mean time, the energy spectrum of the EC decay should
also be convoluted with the Gaussian energy resolution. So we
rewrite Eq. (3) as
\begin{eqnarray}
{{\rm d} \lambda^{}_{\rm EC}\over {\rm d} T^{}} \hspace{-0.2cm} & =
& \hspace{-0.2cm} \int_0^{Q - {\rm min}(m^{}_k)} {\rm d} T^{}_{\rm
c} \, \left[ {G^{2}_{\beta} \over {4 \pi^2}} \, R(T, \, T^{}_{\rm
c}) \, (Q-T^{}_{\rm c}) \right .
\nonumber \\
&& \hspace{-0.2cm} \left . \times \sum_k
|V^{}_{ek}|^2\sqrt{(Q-T^{}_{\rm c})^2-m_k^2}~ \Theta(Q - T^{}_{\rm
c} - m^{}_k) \right .
\nonumber \\
&& \hspace{-0.2cm} \left . \times \sum_i n^{}_i \, C^{}_i \,
\beta_i^2 \, B^{}_i \, {\Gamma^{}_i \over 2\pi} \cdot {1 \over
(T^{}_{\rm c}-E^{}_i)^2+\Gamma_i^2/4} \right] \; .
\end{eqnarray}
The above two equations allow us to calculate the rate of the relic
antineutrino capture on the EC-decaying $^{163}{\rm Ho}$ nuclei
against the corresponding background (i.e., the EC decay itself) in
the presence of flavor effects, which are characterized by both the
neutrino masses $m^{}_k$ and the neutrino mixing matrix elements
$|V^{}_{ek}|$ (for $k=1,2,3$). \\

We proceed to evaluate the flavor effects on both the relic
antineutrino capture rate and the corresponding EC decay rate with
the help of Eqs. (8) and (9). In our numerical calculations we take
$\Delta m^2_{21} \approx 7.6 \times 10^{-5} ~{\rm eV}^2$ and
$|\Delta m^2_{31}| \approx 2.4 \times 10^{-3} ~{\rm eV}^2$ together
with $\theta^{}_{12} \approx 34^\circ$ as typical inputs \cite{GG}.
The impact of $\theta^{}_{13}$ on $\lambda^{}_{\overline{\nu}}$ and
$\lambda^{}_{\rm EC}$ can be examined by allowing its value to vary
within $0^\circ \leq \theta^{}_{13} < 12^\circ$ \cite{GG}. Depending
on the sign of $\Delta m^2_{31}$, two mass patterns of three active
neutrinos are possible: one is the normal mass ordering with $m_1 <
m_2 = \sqrt{m^2_1 + \Delta m^2_{21}} < m_3 = \sqrt{m^2_1 + |\Delta
m_{31}^2|} \hspace{0.05cm}$; and the other is the inverted mass
ordering with $m^{}_3 < m_1 = \sqrt{m_3^2+|\Delta m_{31}^2|} < m_2 =
\sqrt{m_3^2 + |\Delta m_{31}^2| + \Delta m_{21}^2} \hspace{0.05cm}$.
In either case the absolute neutrino mass scale ($m^{}_1$ or
$m^{}_3$) is unknown, but its upper bound is expected to be of
${\cal O}(0.1)$ eV up to ${\cal O}(1)$ eV as constrained by current
experimental and cosmological data \cite{PDG,WMAP10}. As for the
values of those parameters relevant to the atomic levels of
$^{163}{\rm Dy}$, we refer the reader to Ref. \cite{Lusignoli} and
references therein. The energy levels of the captured electrons are
assumed to be fully occupied (i.e., $n^{}_i =1$) and their binding
energies and widths can be found in Table 1 of Ref.
\cite{Lusignoli}. The atomic corrections for the electron exchange
and overlap are neglected (i.e., $B_{i}\simeq 1$), and the ratios of
the squared wave functions at the origin (i.e.,
$\beta^{2}_{i}/\beta^2_{\rm M^{}_1}$) can be found in Table 2 of
Ref. \cite{Lusignoli}. Because the nuclear shape factors $C^{}_i$
are approximately identical in an allowed transition
\cite{Bambynek}, they can be factored out from the sum in Eqs. (3),
(4), (6), (8) and (9). Finally, one should note that the numerical
results of $\lambda^{}_{\overline{\nu}}$ and $\lambda^{}_{\rm EC}$
can be properly normalized by means of the half-life of $^{163}{\rm
Ho}$ through the relation $\lambda^{}_{\rm EC}T^{}_{1/2}=\ln 2$,
where $T^{}_{1/2} \simeq 4570$ yr. Then the distributions of the
number of capture events and the number of background events are
expressed, respectively, as
\begin{eqnarray}
\frac{{\rm d} N^{}_{\rm S}}{{\rm d} T} \hspace{-0.2cm} & = &
\hspace{-0.2cm} {1 \over \lambda^{}_{\rm EC}} \cdot {{\rm d}
\lambda^{}_{\overline{\nu}} \over {\rm d} T} \cdot {\ln 2 \over
T^{}_{1/2}} \, N^{}_{\rm T} \, t \; ,
\nonumber \\
\frac{{\rm d} N^{}_{\rm B}}{{\rm d} T} \hspace{-0.2cm} & = &
\hspace{-0.2cm} {1 \over \lambda^{}_{\rm EC}} \cdot {{\rm d}
\lambda^{}_{\rm EC}\over {\rm d} T} \cdot {\ln 2 \over T^{}_{1/2}}
\, N^{}_{\rm T} \, t \;
\end{eqnarray}
for a given target factor $N^{}_{\rm T}$ (i.e., the number of
$^{163}{\rm Ho}$ atoms of the target) and for a given exposure time
$t^{}_{}$ in the experiment.

Let us first look at the energy spectrum of the EC decay of
$^{163}{\rm Ho}$. Our main concern is the flavor effects on the
ratios of the peak heights and the fine structure near the endpoint
of the energy spectrum. The numerical results are shown in Fig. 1,
where $\theta^{}_{13} = 10^\circ$ has typically been input. Fig.
1(a) and Fig. 1(b) show that two ratios of the peak heights are
completely insensitive to the lightest neutrino mass $m^{}_0$, which
is identified with $m^{}_1$ for the normal mass hierarchy (i.e.,
$m^{}_1 < m^{}_2 < m^{}_3$) or with $m^{}_{3}$ for the inverted mass
hierarchy (i.e., $m^{}_3 < m^{}_1 < m^{}_2$). Given $m^{}_0 \leq
0.5$ eV, the actual variations of two ratios are only at the ${\cal
O}(10^{-6})$ level. Hence it is hopeless to obtain any information
about the absolute mass scale of three neutrinos or their mass
hierarchy by measuring these ratios. However, it seems very
promising to acquire knowledge of the $Q$-value from these ratios,
as one can clearly see in Fig. 1(c) and Fig. 1(d). In Fig. 1(e) and
Fig. 1(f) we illustrate how the ratios of the peak heights change
with the energy resolution $\Delta$ for given values of $m^{}_0$ and
$Q$. We see that the changes are at the percent level if $\Delta
\leq 1$ eV is taken.

The fine structure near the spectral endpoint of the EC decay of
$^{163}{\rm Ho}$ is illustrated in Fig. 2, where $\theta^{}_{13} =
10^\circ$ has also been input. The location of the endpoint is
essentially determined by the smallest neutrino mass. It might be
possible to distinguish between the normal and inverted neutrino
mass hierarchies, since they have different effects on the energy
spectrum near its endpoint. Especially for the inverted mass
hierarchy, an obvious distortion occurs at the position $T^{}_{\rm
c}\simeq Q- m^{}_1$ with $m^{}_1 = \sqrt{m^{2}_{3}+|\Delta
m^{2}_{31}|} \hspace{0.05cm}$ due to the fact that the lightest mass
eigenstate $\nu^{}_{3}$ hidden in $\nu^{}_{e}$ is associated with
the smallest neutrino mixing matrix element $|V^{}_{e3}|^{2}
=\sin^2\theta^{}_{13}$. In the region of $T^{}_{\rm c}> Q-m^{}_1$,
only the $\nu^{}_{3}$ component contributes to the spectrum. In
comparison, the contributions of $\nu^{}_1$ and $\nu^{}_2$
components to the spectrum are dominant in the region of $T^{}_{\rm
c}< Q-m^{}_1$. One might therefore be able to determine the
magnitude of $|V^{}_{e3}|^{2}$ if the endpoint distortion in the
energy spectrum could really be observed. A similar distortion
cannot appear in the case of the normal neutrino mass hierarchy,
simply because the contribution of the lightest mass eigenstate
$\nu^{}_{1}$ to the spectrum is of the same order of
magnitude as those of $\nu^{}_2$ and $\nu^{}_3$. \\

Let us now focus on the relic antineutrino capture on $^{163}{\rm
Ho}$ nuclei against the EC decay background by taking account of the
flavor effects and examining the $Q$-value dependence. In our
calculations we assume 30 kg $^{163}{\rm Ho}$ as a reference isotope
source. The numerical results are presented in Fig. 3, Fig.4 and
Fig. 5, where the relic antineutrino capture rate is illustrated as
a function of the overall energy release $T$ in a possible
experiment. Some comments and discussions are in order.

(1) Given $\theta^{}_{13} = 10^\circ$ as a typical input, Fig. 3
shows the effects of the neutrino mass hierarchy on the
C$\overline{\nu}$B signature and the corresponding background. The
value of the finite energy resolution $\Delta$ is chosen in such a
way that only a single peak of the signature can be seen; namely, at
least one peak of the signature is not allowed to fall into the
shade of the background. Eq. (8) tells us that the contribution of
each antineutrino mass eigenstate $\overline{\nu}^{}_k$ (for
$k=1,2,3$) to the capture rate is located at $T = T^{}_k \equiv
E^{}_{\overline{\nu}^{}_k}+Q$ and weighted by the relic antineutrino
number density $n^{}_{\overline{\nu}^{}_k} \simeq 56 ~{\rm cm}^{-3}$
and the flavor mixing matrix element $|V^{}_{ek}|^{2}$. On the other
hand, Eq. (9) tells us that the energy spectrum of the EC decay near
its endpoint is dominated by the lightest neutrino mass eigenstate
hidden in $\overline{\nu}^{}_{e}$ and sensitive to the energy
resolution $\Delta$. As the smallest neutrino mass ($m^{}_1$ in the
left panel of Fig. 3 or $m^{}_3$ in the right panel of Fig. 3)
increases from 0 to 0.1 eV, the C$\overline{\nu}$B signature moves
towards the larger $T-Q$ region. In comparison, the shift of the
corresponding background is less obvious because the smallest
neutrino mass and $\Delta$ have the opposite effects on the location
of the spectral endpoint of the EC decay. Hence the distance between
the peak of the signature and the background becomes larger for a
larger value of the smallest neutrino mass, and accordingly the
required energy resolution $\Delta$ becomes less stringent.
Comparing between Fig. 3(c) and Fig. 3(d), for example, one can also
see that it is more or less easier to detect the C$\overline{\nu}$B
in the $\Delta m^2_{31} <0$ case, where the signature is separated
more obviously from the background. The reason is simply that the
dominant antineutrino mass eigenstates $\overline{\nu}^{}_1$ and
$\overline{\nu}^{}_2$ have slightly larger eigenvalues in this case
than in the $\Delta m^2_{31} > 0$ case. Note that more than one peak
may emerge in the distribution of the relic antineutrino capture
rate, as shown in Fig. 3(a) and Fig. 3(b), provided the energy
resolution is sufficiently good. But the requirement $\Delta \leq
m^{}_{k}/2$ \cite{Vogel} is in general needed so as to resolve the
corresponding peak.

(2) The effect of the unknown neutrino mixing angle $\theta^{}_{13}$
on the C$\overline{\nu}$B detection is examined in Fig. 4, where
both the $\Delta m^2_{31}>0$ case with $m^{}_1 =0$ and the $\Delta
m^2_{31}<0$ case with $m^{}_3 =0$ are taken into account. The energy
resolution $\Delta$ is chosen in such a way that the signature in
Fig. 4(b) with $m^{}_3 =0$ and $\theta^{}_{13}=0^\circ$ should be
clearly seen. In the $\Delta m^2_{31}>0$ case, Fig. 4(a), Fig. 4(c)
and Fig. 4(e) show that a change of $\theta_{13}$ in its allowed
region (i.e., $\theta^{}_{13} < 12^\circ$) almost has no effect on
both the signature and the background. In the $\Delta m^2_{31}<0$
case, however, the background turns out to be sensitive to a change
of $\theta^{}_{13}$ as one can see from Fig. 4(b), Fig. 4(d) and
Fig. 4(f). As $\theta^{}_{13}$ becomes larger, the signature
essentially keeps intact but the background apparently moves towards
the larger $T-Q$ region, making the signature partly fall into the
shade. Note that the distribution of the C$\overline{\nu}$B capture
rate may have a slight distortion only when $\theta^{}_{13}$ is
sufficiently large, as shown in Fig. 4(e) and Fig. 4(f). Hence it is
in practice almost impossible to probe the magnitude of
$\theta^{}_{13}$ by detecting the C$\overline{\nu}$B.

(3) The $Q$-value may sensitively affect the absolute capture rate
of relic antineutrinos, but it does not influence the relative
locations of the signature and its corresponding background. This
property is illustrated in Fig. 5, where the smallest neutrino mass
(i.e., $m^{}_1$ in the $\Delta m^2_{31}>0$ case or $m^{}_3$ in the
$\Delta m^2_{31}<0$ case) has been taken to be zero for simplicity.
Allowing $Q$ to vary from 2.3 keV to 2.8 keV, we find that the
absolute capture rate undergoes a decrease of as much as one order
of magnitude. The reason for this feature can be explained with the
help of Eq. (8). The binding energies $E^{}_i$ of the seven
available atomic levels are all smaller than 2.1 keV, and thus the
individual capture rates at $T^{}_k \equiv
E^{}_{\overline{\nu}^{}_k}+Q$ are strongly suppressed by the
corresponding Breit-Wigner distributions. A larger $Q$-value implies
a stronger suppression of the overall capture rate
$\lambda^{}_{\overline{\nu}}$, leading to a worse signature of the
C$\overline{\nu}$B in the experiment. That is why the isotope of a
smaller $Q$-value is eagerly wanted \cite{Lusignoli}, in order to
make this interesting C$\overline{\nu}$B detection
method more feasible in the foreseeable future. \\

In summary, we have examined the flavor effects on a possible
detection of the C$\overline{\nu}$B by using a target made of
$^{163}{\rm Ho}$ nuclei. On the one hand, we have discussed the
energy spectrum of the EC decay of $^{163}{\rm Ho}$ by looking at
how sensitive the ratios of the peak heights and the fine structure
near the spectral endpoint are to the neutrino mass hierarchy. On
the other hand, we have presented a detailed analysis of the effects
of different neutrino masses and reasonable values of the smallest
neutrino mixing angle $\theta^{}_{13}$ on the relic antineutrino
capture rate and the corresponding EC decay background. Our results
demonstrate that such flavor effects, which were not considered in
Ref. \cite{Lusignoli}, can be appreciable and even important in some
cases.

At least three factors may apparently affect the observability of
relic antineutrinos in such a calorimetric experiment. The first one
is the number of the target particles $N^{}_{\rm T}$, which is the
only adjustable parameter for a given measurement to increase the
total capture events. In our numerical analysis we have taken 30 kg
$^{163}{\rm Ho}$ as a reference isotope source, just for the purpose
of illustration. A much larger target is certainly welcome, but it
will certainly be a bigger challenge to a realistic experiment. The
second factor is the finite energy resolution $\Delta$, which has a
crucial impact on the signature-to-background ratio. As the
precision of $\Delta$ is gradually improved, it should be possible
to establish a signature of relic antineutrinos just beyond the
spectral endpoint of the EC decay of $^{163}{\rm Ho}$ in the future.
We admit that $\Delta = 0.015$ eV taken in our numerical examples is
too optimistic, because it is about two orders of magnitude better
than the present achievement. If the energy resolution were really
perfect, we would even be able to resolve the multiple peaks coming
from different neutrino mass eigenstates and then obtain some
information on neutrino masses and flavor mixing angles. But it will
be a great success even if only a single peak of the signature can
in practice be resolved. The third factor is the number density of
relic antineutrinos $n^{}_{\overline{\nu}^{}_k}$ (for $k=1,2,3$)
around the Earth. Its value might be much larger than
$n^{}_{\nu^{}_k} = n^{}_{\overline{\nu}^{}_k} \simeq 56 ~ {\rm
cm}^{-3}$ predicted by the standard Big Bang cosmology, if the
gravitational clustering effect is significant \cite{Wong}. A larger
value of $n^{}_{\overline{\nu}^{}_k}$ will always be a good news for
the detection of the C$\overline{\nu}$B.

To conclude, the C$\nu$B and C$\overline{\nu}$B can in principle be
detected by means of the $\beta$-decaying nuclei (e.g., $^3{\rm H}$)
and the EC-decaying nuclei (e.g., $^{163}{\rm Ho}$), respectively.
Although the present experimental techniques are unable to lead us
to a guaranteed measurement of relic neutrinos and antineutrinos in
the near future, we might be able to make a success of this great
exploration in the long term. Of course, it would be much harder to
identify flavor effects in the signatures of relic neutrinos and
antineutrinos. Let us reiterate that a direct laboratory measurement
of the C$\nu$B and C$\overline{\nu}$B is extremely important and
thus deserves more attention and our best efforts. At least from a
historical point of view, nothing is completely hopeless in neutrino
physics. \\

This work was supported in part by the China Postdoctoral Science
Foundation under grant No. 20100480025 (Y.F.L.) and in part by the
National Natural Science Foundation of China under grant No.
10875131 (Z.Z.X.).

\newpage

\begin{figure}[p!]
\begin{center}
\begin{tabular}{cc}
\includegraphics*[bb=18 18 280 216, width=0.46\textwidth]{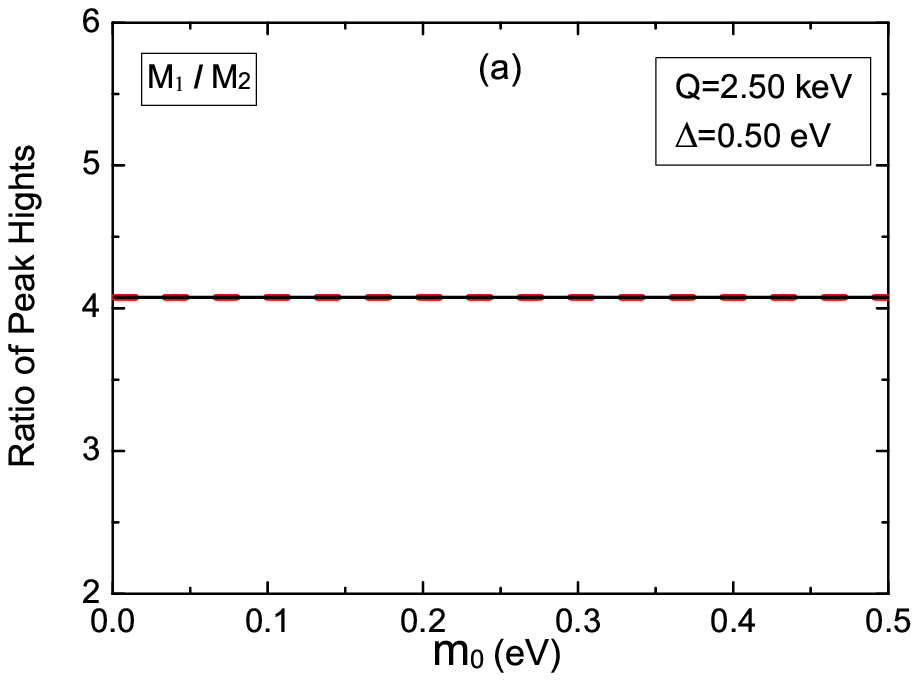}
&
\includegraphics*[bb=18 18 280 216, width=0.46\textwidth]{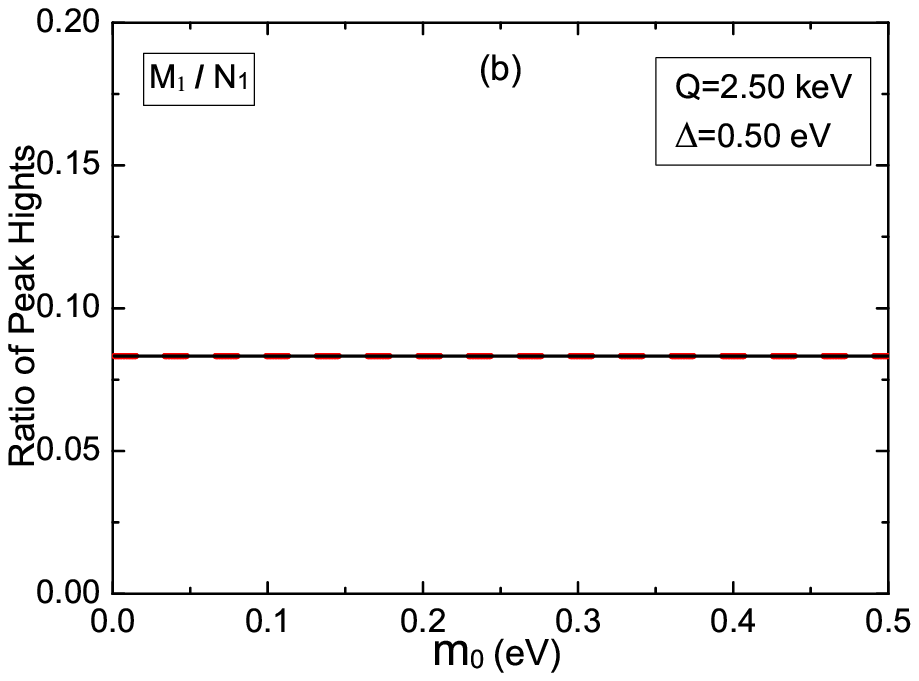}
\\
\includegraphics*[bb=18 18 280 216, width=0.46\textwidth]{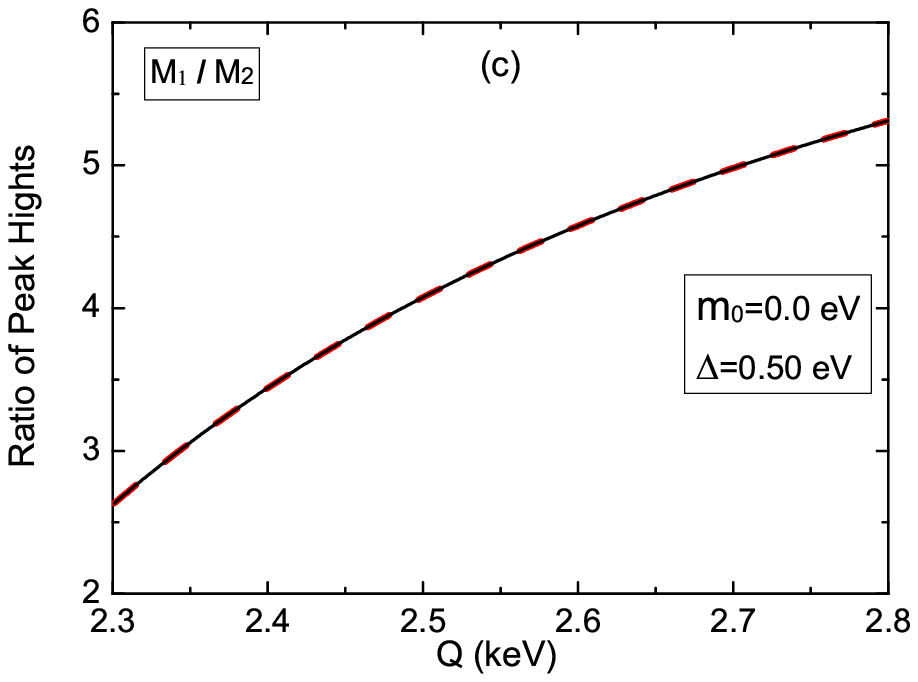}
&
\includegraphics*[bb=18 18 280 216, width=0.46\textwidth]{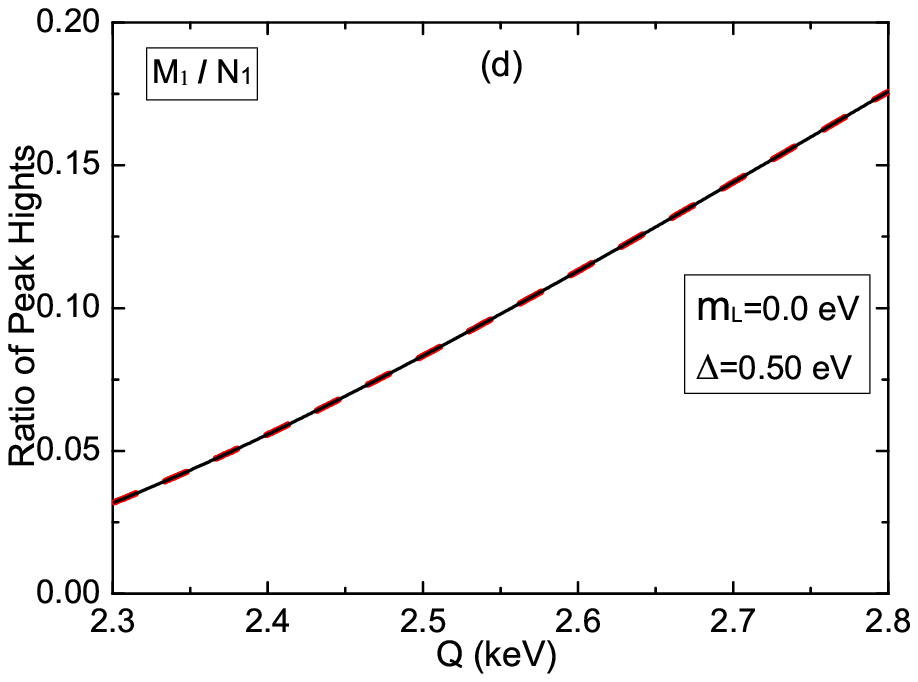}
\\
\includegraphics*[bb=18 18 280 216, width=0.46\textwidth]{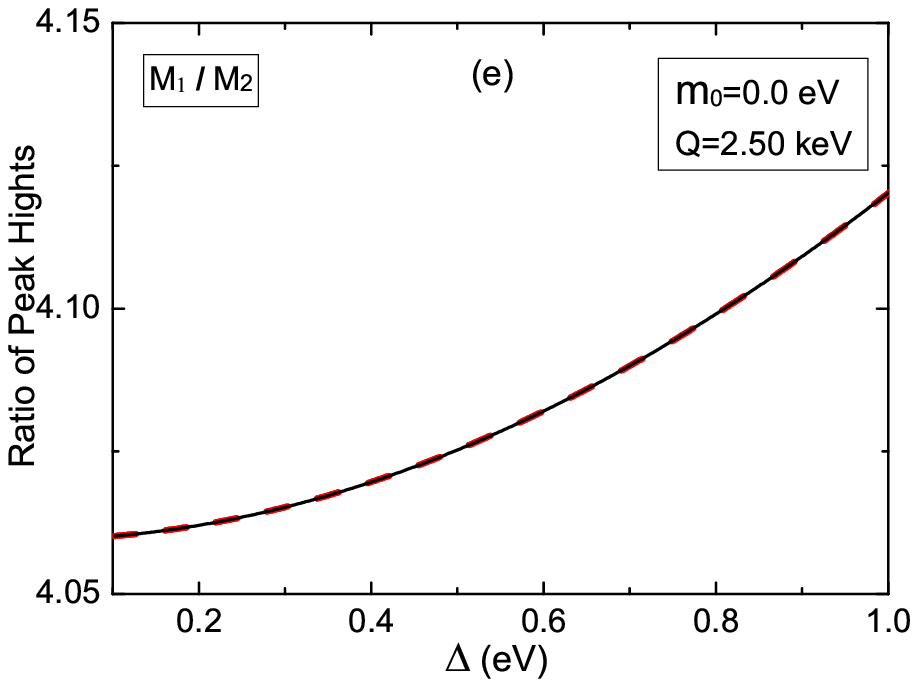}
&
\includegraphics*[bb=18 18 280 216, width=0.46\textwidth]{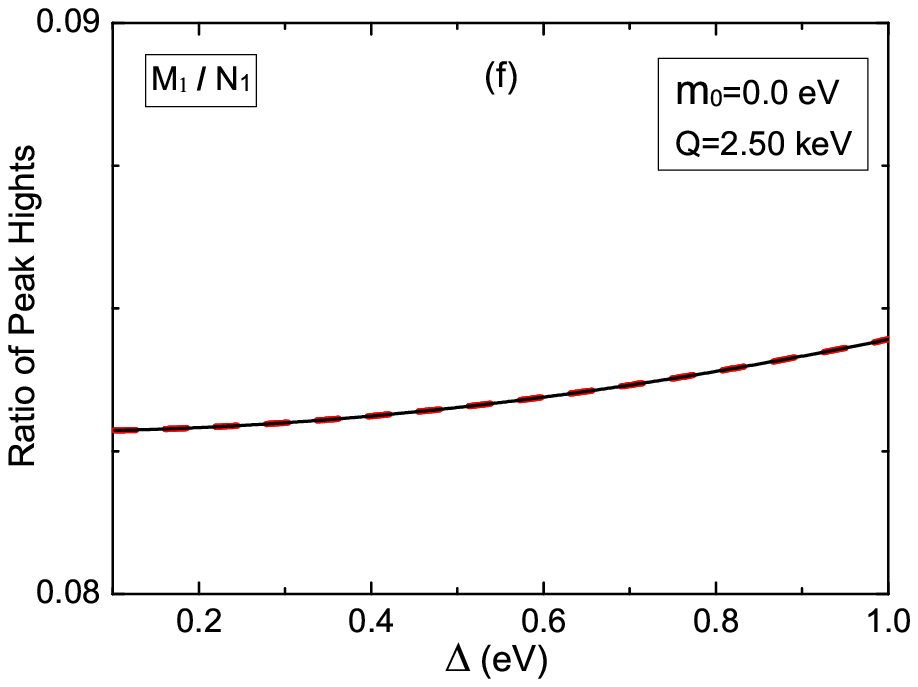}
\end{tabular}
\end{center}
\caption{The energy spectrum of the EC decay of $^{163}{\rm Ho}$:
the ratios of the peak heights $M^{}_1/M^{}_2$ and $M^{}_1/N^{}_1$
changing with the lightest neutrino mass
$m^{}_0$ (equal to $m^{}_1$ in the $m^{2}_{31}>0$ case with a solid
curve or to $m^{}_3$ in the $m^{2}_{31}<0$ case with a dashed curve),
the $Q$-value and the energy resolution $\Delta$. Here
$\theta^{}_{13} =10^\circ$ has typically been input.}
\end{figure}

\begin{figure}[p!]
\begin{center}
\begin{tabular}{cc}
\includegraphics*[bb=16 18 278 220, width=0.6\textwidth]{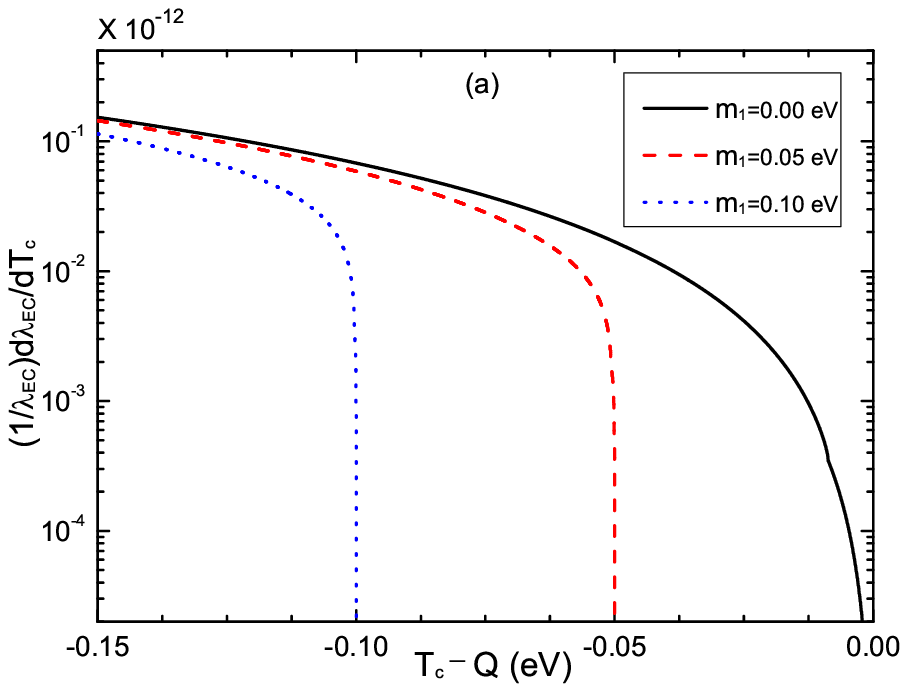}
\\
\includegraphics*[bb=16 18 278 220, width=0.6\textwidth]{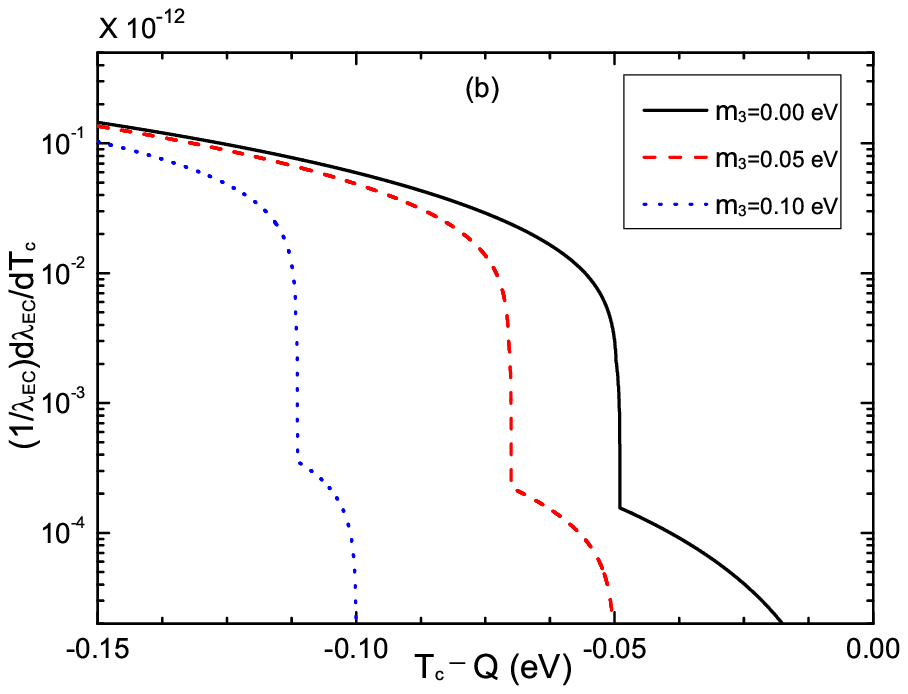}
\end{tabular}
\end{center}
\caption{The energy spectrum of the EC decay of $^{163}{\rm Ho}$:
the fine structure near its endpoint in the $m^{2}_{31}>0$ (top panel)
or $m^{2}_{31}<0$ (bottom panel) case. Here $Q=2.5$ keV and
$\theta^{}_{13} =10^\circ$ have typically been input.}
\end{figure}

\begin{figure}[p!]
\begin{center}
\begin{tabular}{cc}
\includegraphics*[bb=18 18 276 236, width=0.46\textwidth]{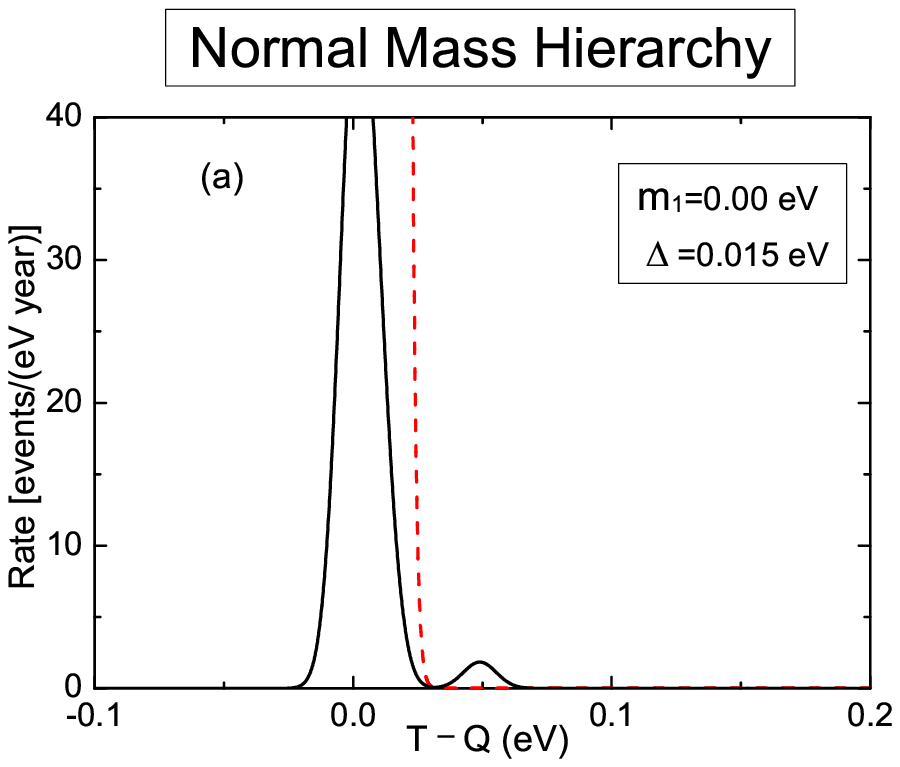}
&
\includegraphics*[bb=18 18 276 236, width=0.46\textwidth]{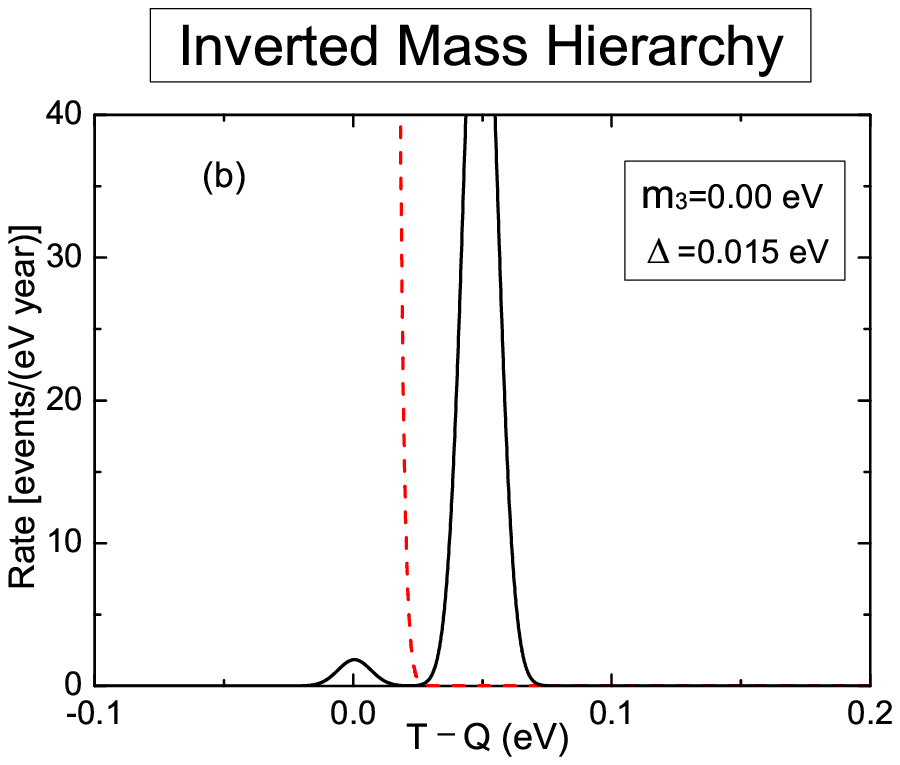}
\\
\includegraphics*[bb=18 18 276 210, width=0.46\textwidth]{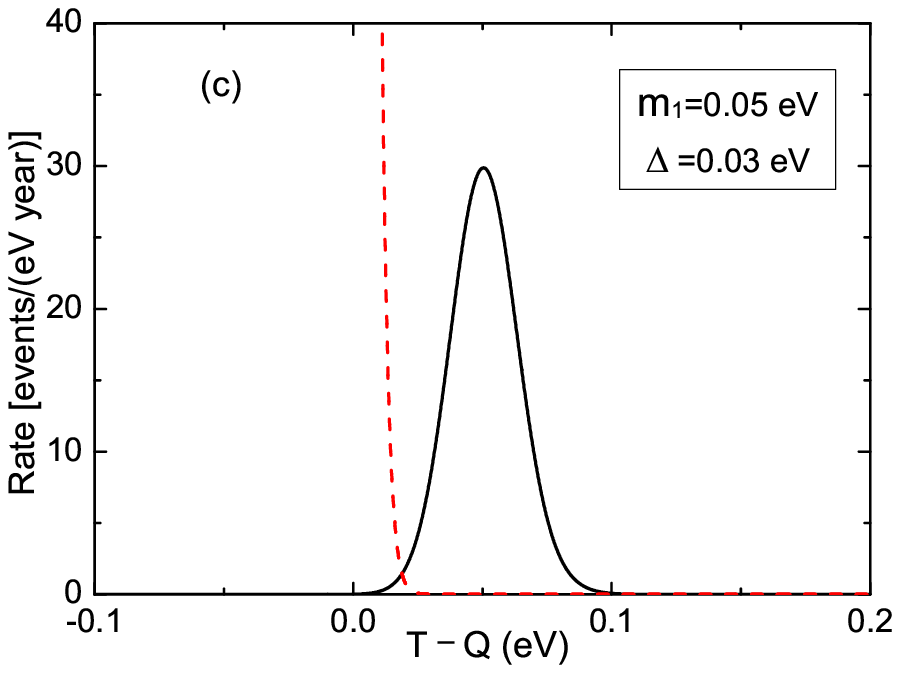}
&
\includegraphics*[bb=18 18 276 210, width=0.46\textwidth]{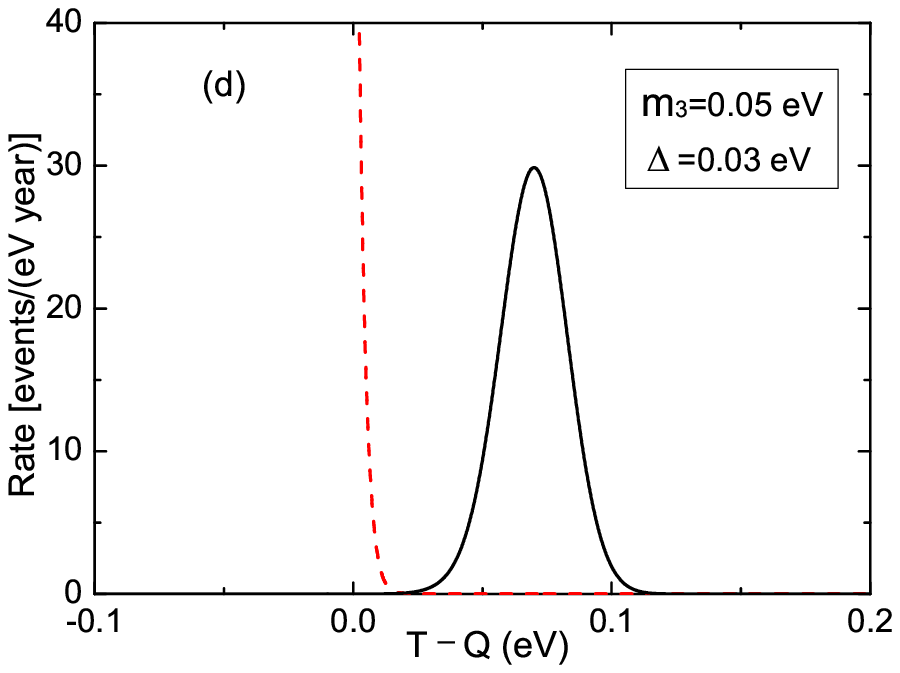}
\\
\includegraphics*[bb=18 18 276 210, width=0.46\textwidth]{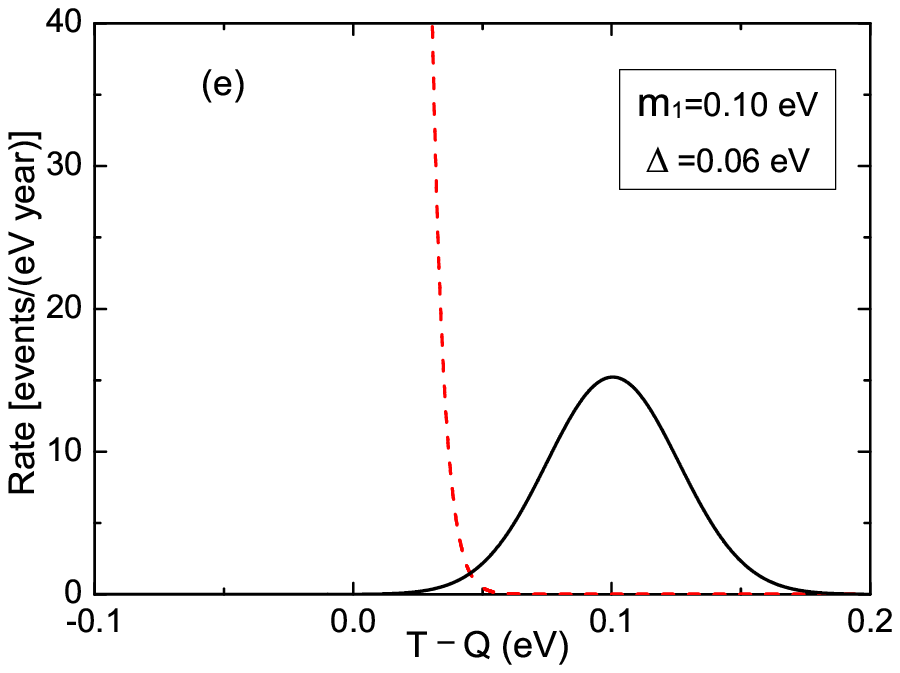}
&
\includegraphics*[bb=18 18 276 210, width=0.46\textwidth]{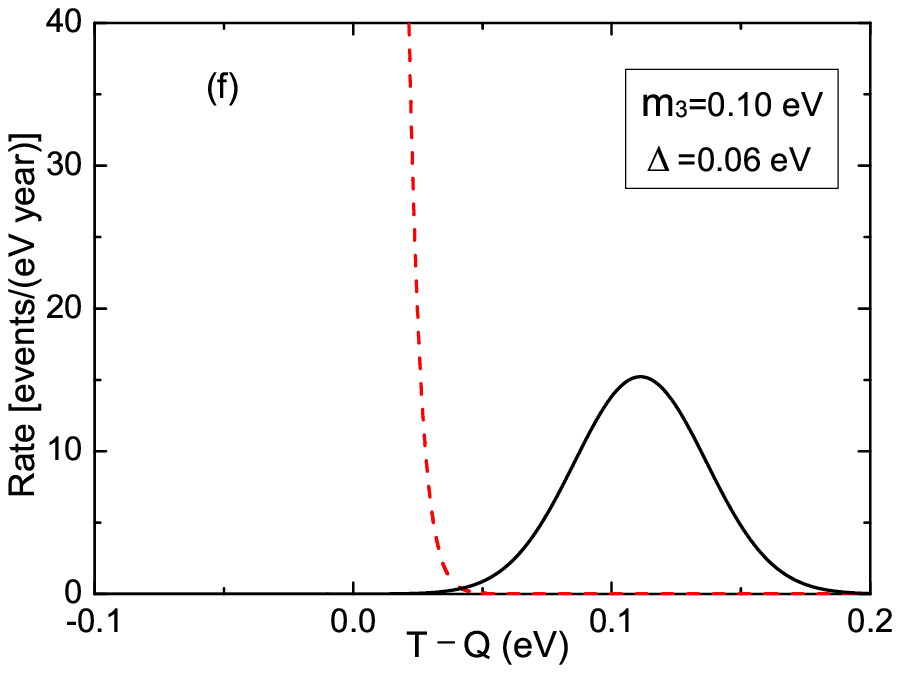}
\end{tabular}
\end{center}
\caption{Effects of neutrino masses on the relic antineutrino
capture rate as a function of the overall energy release $T$ in the
$\Delta m^2_{31} >0$ (left panel) or $\Delta m^2_{31}<0$ (right
panel) case. The solid and dashed curves represent the
C$\overline{\nu}$B signature and its background, respectively. The
value of the finite energy resolution $\Delta$ is chosen in such a
way that only a single peak of the signature can be seen. Here
$\theta^{}_{13} =10^\circ$ has typically been input.}
\end{figure}

\begin{figure}[p!]
\begin{center}
\begin{tabular}{cc}
\includegraphics*[bb=18 18 276 236, width=0.46\textwidth]{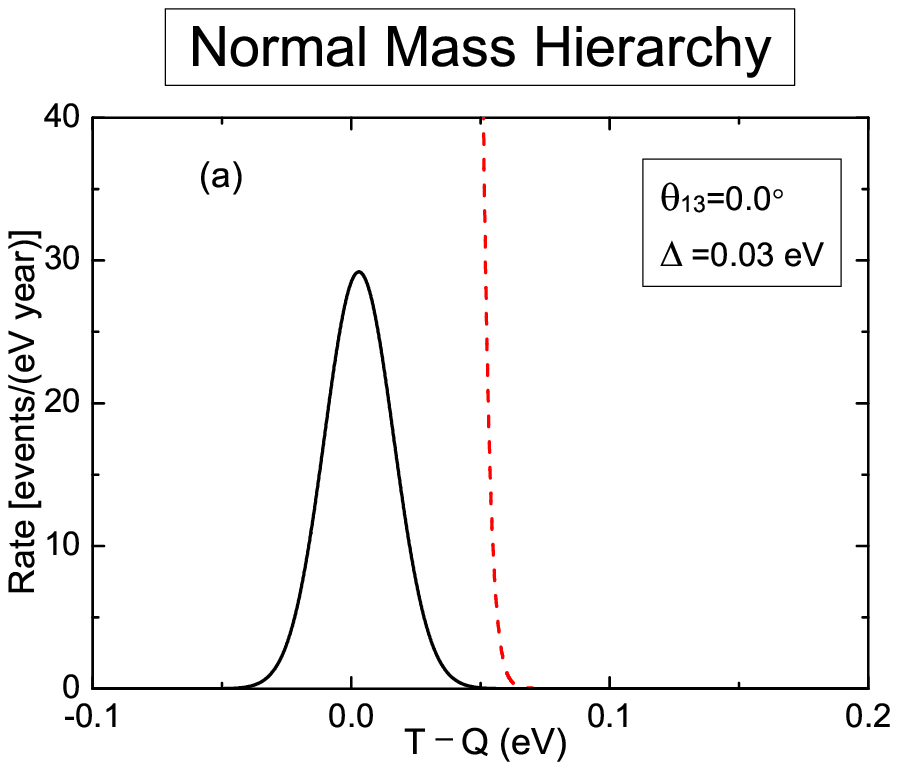}
&
\includegraphics*[bb=18 18 276 236, width=0.46\textwidth]{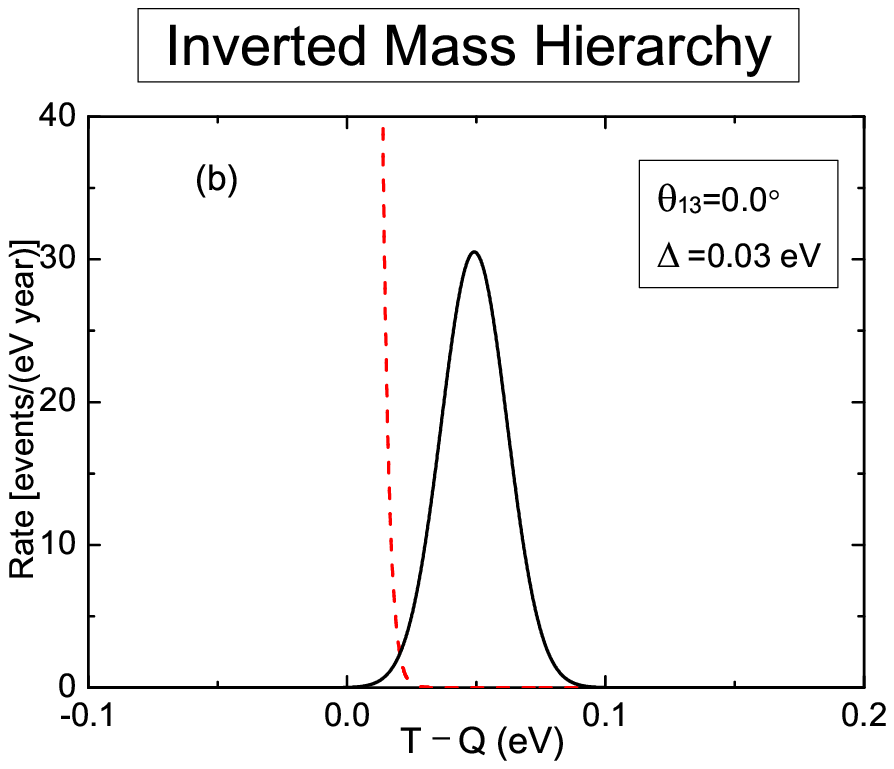}
\\
\includegraphics*[bb=18 18 276 210, width=0.46\textwidth]{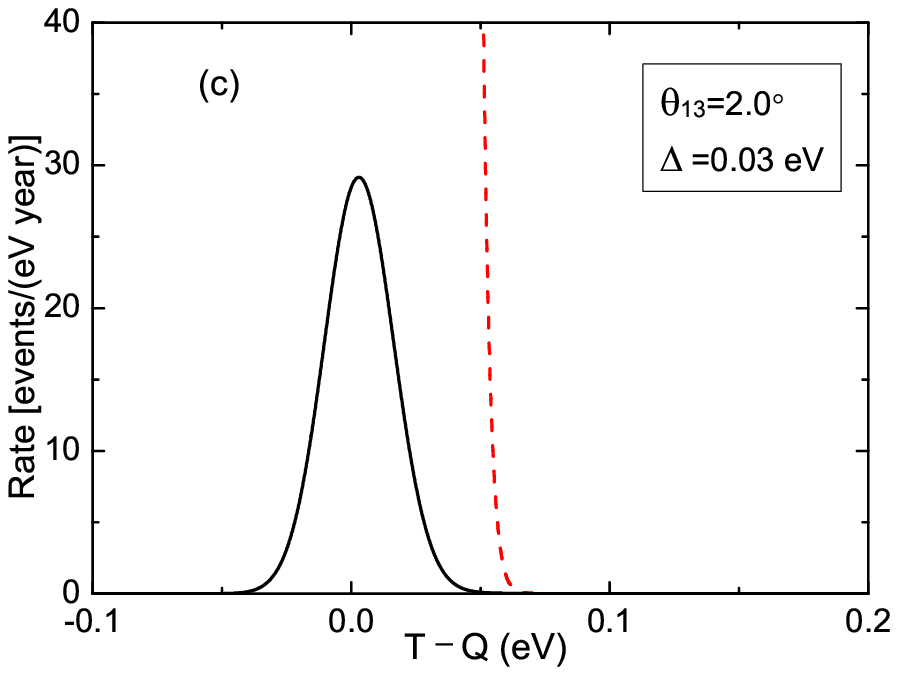}
&
\includegraphics*[bb=18 18 276 210, width=0.46\textwidth]{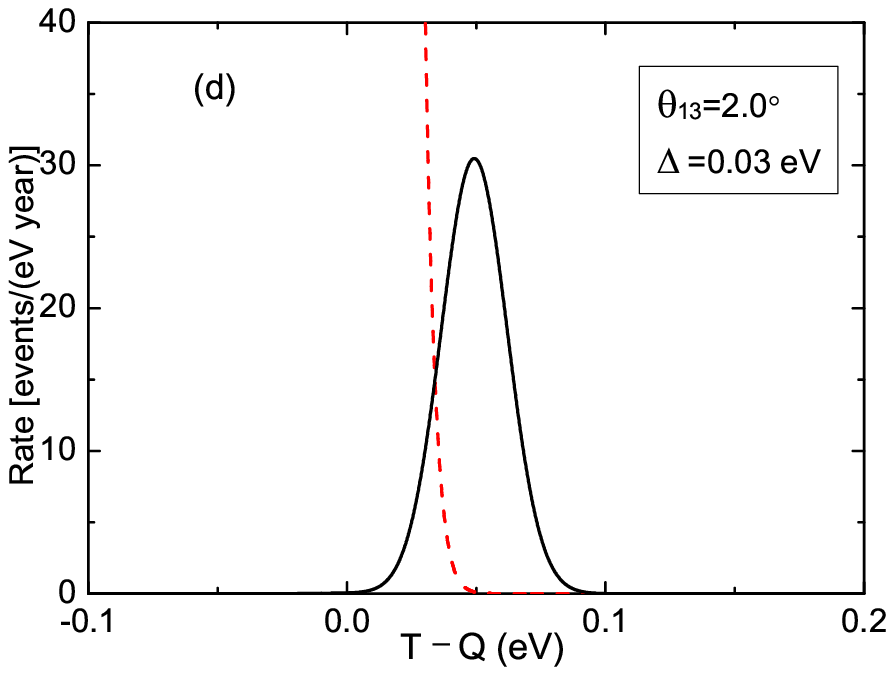}
\\
\includegraphics*[bb=18 18 276 210, width=0.46\textwidth]{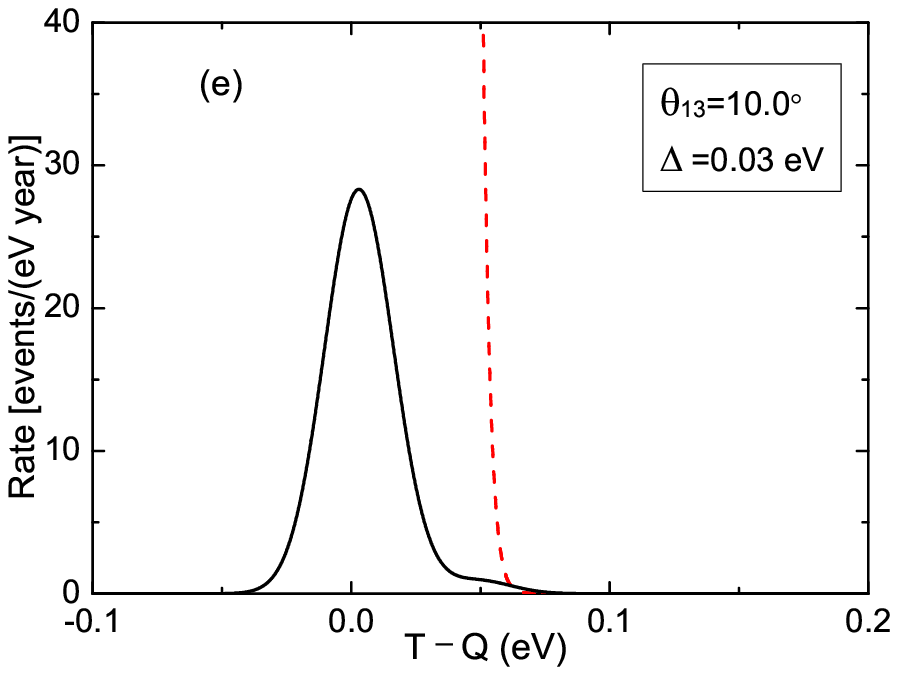}
&
\includegraphics*[bb=18 18 276 210, width=0.46\textwidth]{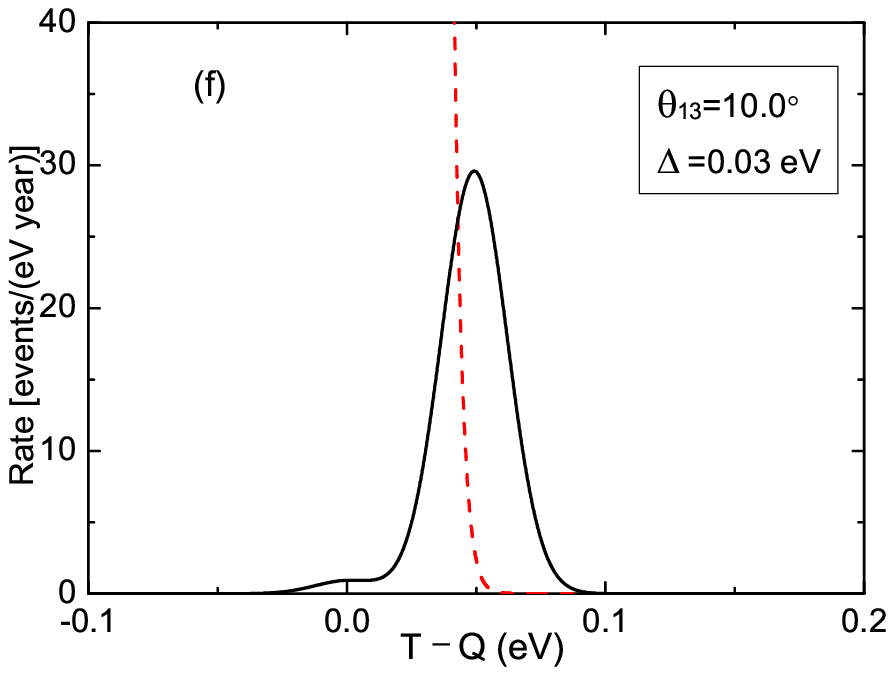}
\end{tabular}
\end{center}
\caption{Effects of $\theta^{}_{13}$ on the relic antineutrino
capture rate as a function of the overall energy release $T$ in the
$\Delta m^2_{31} >0$ case with $m^{}_1 =0$ (left panel) or in the
$\Delta m^2_{31}<0$ case with $m^{}_3 =0$ (right panel). The solid
and dashed curves represent the C$\overline{\nu}$B signature and its
background, respectively. The value of the finite energy resolution
$\Delta$ is chosen in such a way that the signature in Fig. 4(b)
with $\theta_{13}=0^\circ$ should be clearly seen.}
\end{figure}

\begin{figure}[p!]
\begin{center}
\begin{tabular}{cc}
\includegraphics*[bb=18 18 278 249, width=0.46\textwidth]{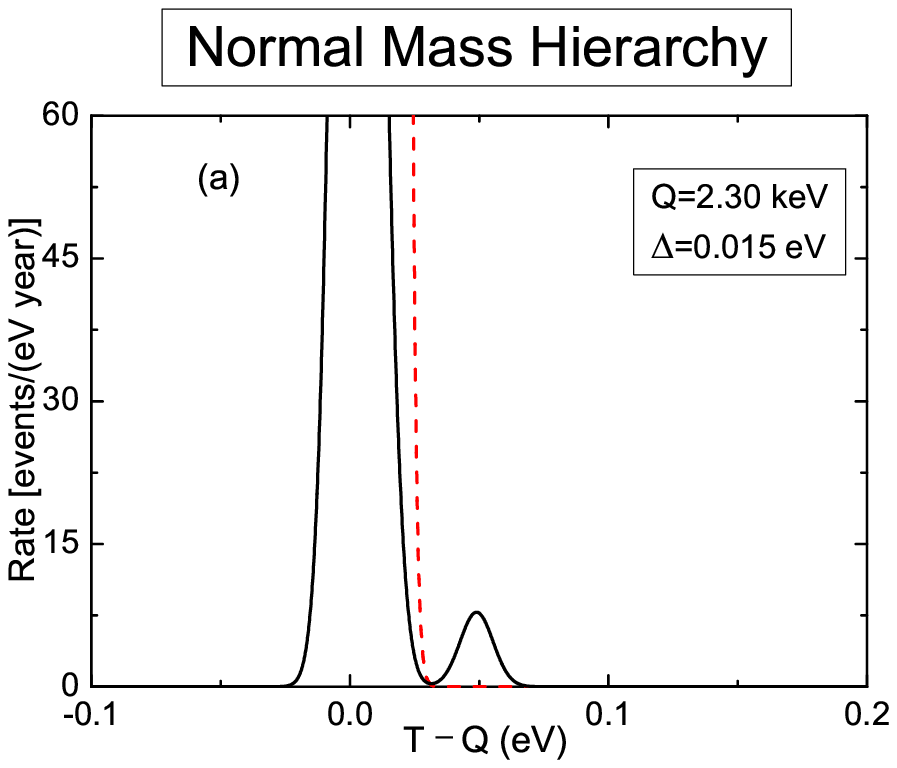}
&
\includegraphics*[bb=18 18 278 249, width=0.46\textwidth]{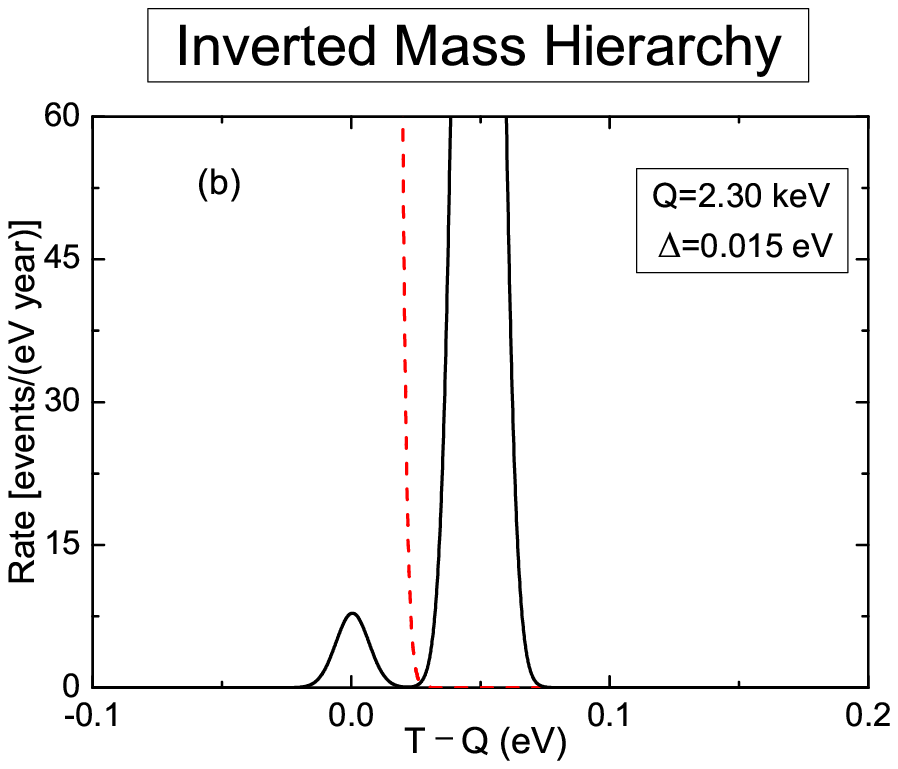}
\\
\includegraphics*[bb=18 18 278 212, width=0.46\textwidth]{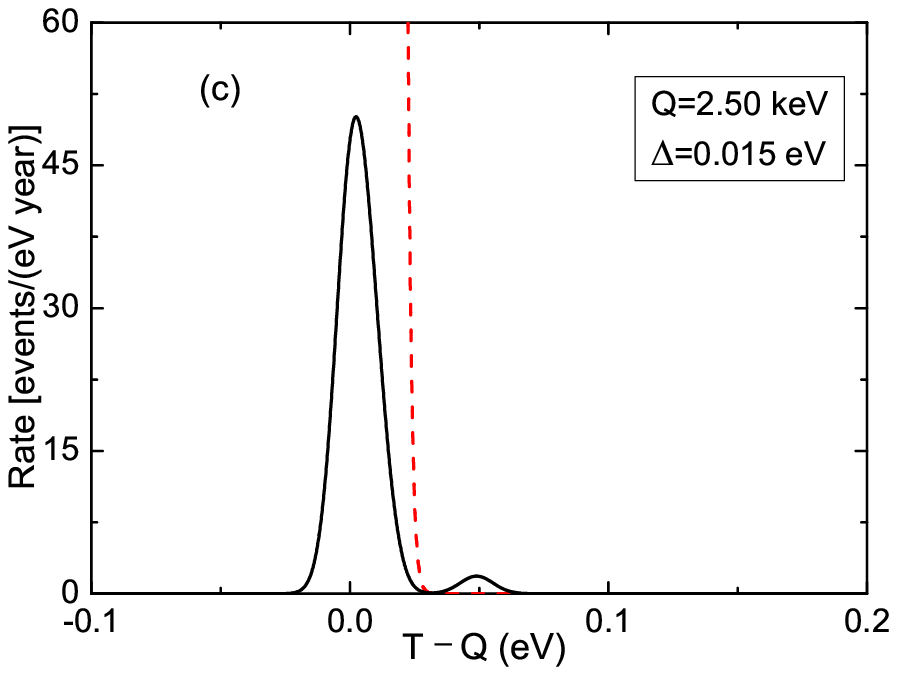}
&
\includegraphics*[bb=18 18 278 212, width=0.46\textwidth]{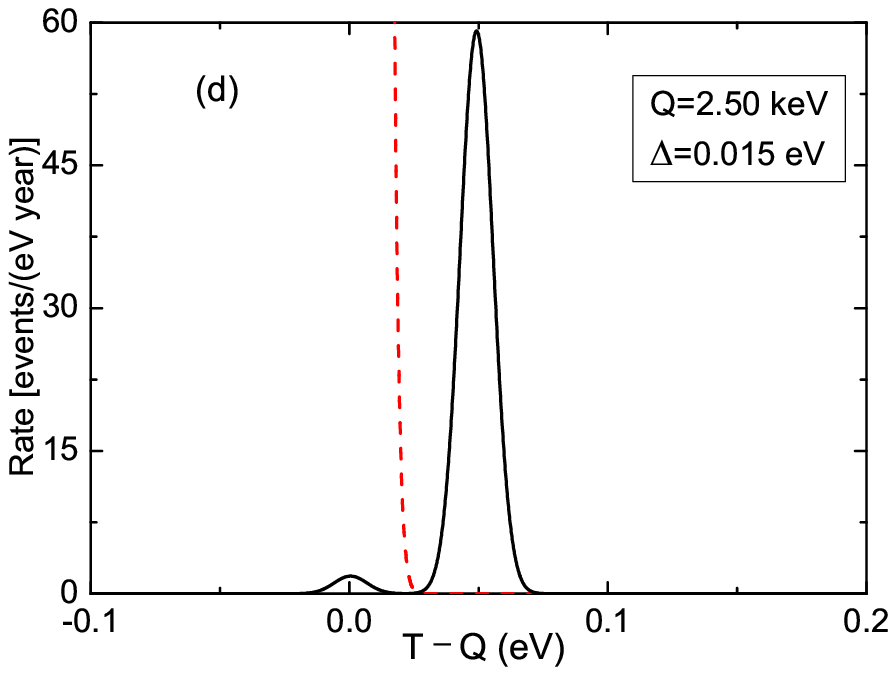}
\\
\includegraphics*[bb=18 18 278 212, width=0.46\textwidth]{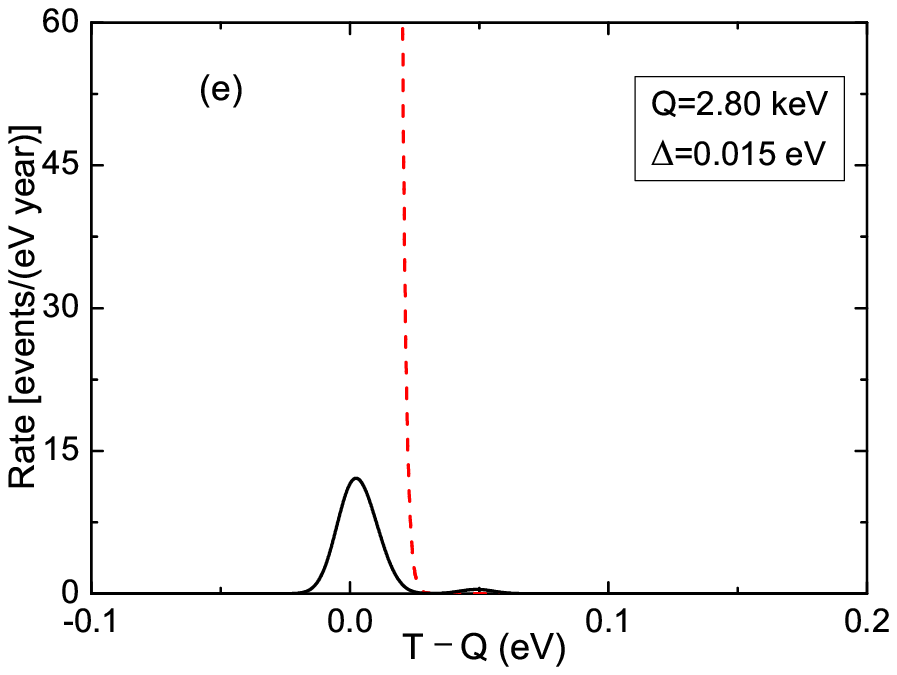}
&
\includegraphics*[bb=18 18 278 212, width=0.46\textwidth]{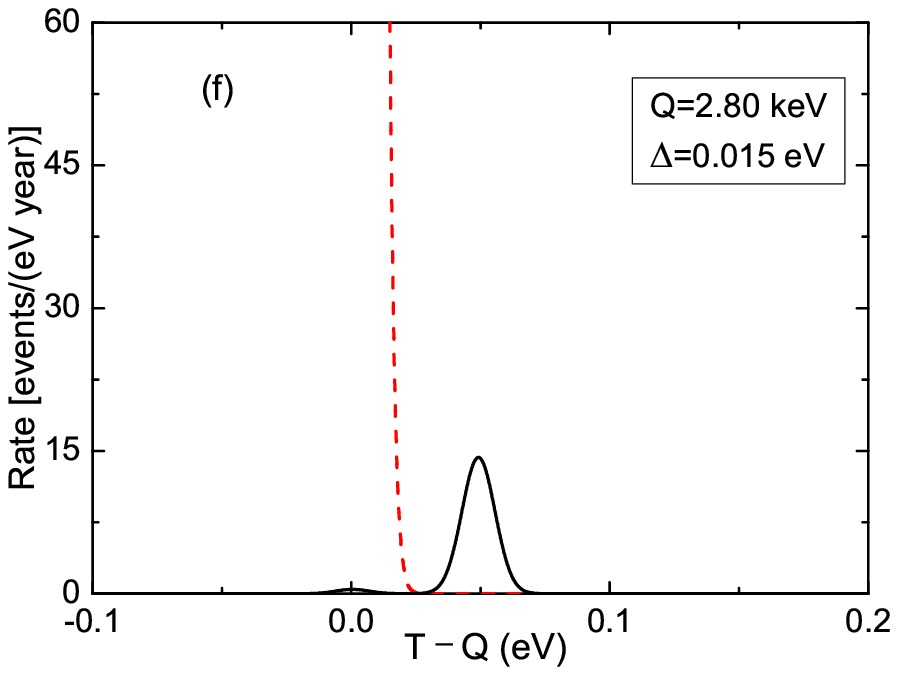}
\end{tabular}
\end{center}
\caption{Effects of the $Q$-value on the relic antineutrino capture
rate as a function of the overall energy release $T$ in the $\Delta
m^2_{31} >0$ case with $m^{}_1 =0$ (left panel) or in the $\Delta
m^2_{31}<0$ case with $m^{}_3 =0$ (right panel). The solid and
dashed curves represent the C$\overline{\nu}$B signature and its
background, respectively. The value of the finite energy resolution
$\Delta$ is chosen in such a way that only a single peak of the
signature can be seen. Here $Q=2.3$ keV, 2.5 keV and 2.8 keV
together with $\theta^{}_{13} =10^\circ$ have typically been input.}
\end{figure}

\end{document}